\documentclass[useAMS,usenatbib,usegraphicx]{mn2e}
\usepackage{amssymb,amsmath,amsfonts,graphicx}
\usepackage{epsfig,color}

\newcommand{\apjl}{ApJL}
\newcommand{\apjs}{ApJS}

\newcommand{\aap}{A\&A}

\usepackage{times}
\usepackage[bookmarks,bookmarksnumbered,colorlinks=true, citecolor=blue, linkcolor=black]{hyperref}

\def\eqsim{\mathrel{\raise0.35ex\hbox{$\scriptstyle =$}\kern-0.6em
\lower0.40ex\hbox{{$\scriptstyle \sim$}}}}
\def\gtrsim{\mathrel{\raise0.35ex\hbox{$\scriptstyle >$}\kern-0.6em
\lower0.40ex\hbox{{$\scriptstyle \sim$}}}}
\def\lesssim{\mathrel{\raise0.35ex\hbox{$\scriptstyle <$}\kern-0.6em
\lower0.40ex\hbox{{$\scriptstyle \sim$}}}}

\title[Low-mass end of the baryonic Tully-Fisher relation]{The low-mass end of the baryonic Tully-Fisher relation}

\author[Sales et al.]{
\parbox[t]{\textwidth}{
Laura V. Sales$^{1}$$\thanks{E-mail: lsales@ucr.edu}$, 
Julio F. Navarro$^{2,3}$,
Kyle Oman$^{2}$,
Azadeh Fattahi$^{2}$,
Ismael Ferrero$^{4,5}$,
Mario Abadi$^{4,5}$,
Richard Bower$^{6}$,
Robert A. Crain$^{7}$,
Carlos S. Frenk$^{6}$,
Till Sawala$^{8}$, 
Matthieu Schaller$^{6}$, 
Joop Schaye$^{9}$,
Tom Theuns$^{6}$ and 
Simon D. M White$^{10}$
} 
\\
\\
  $^{1}$ Department of Physics and Astronomy, University of California, Riverside, CA, 92521, USA\\
  $^{2}$ Department of Physics and Astronomy, University of Victoria, Victoria, BC V8P 5C2, Canada\\
  $^{3}$ Senior CIfAR Fellow\\
  $^{4}$ Observatorio Astron\'omico, Universidad Nacional de C\'ordoba, C\'ordoba, X5000BGR, Argentina\\
  $^{5}$ Instituto de Astronom{\'i}a Te\'orica y Experimental - CONICET, Laprida 922 X5000BGR, Argentina\\
  $^{6}$ Institute for Computational Cosmology, Department of Physics, University of Durham, South Road, Durham DH1 3LE, UK\\
  $^{7}$ Astrophysics Research Institute, Liverpool John Moores University, 146 Brownlow Hill, Liverpool L3 5RF, UK\\
  $^{8}$ Department of Physics, University of Helsinki, Gustaf H\"allstr\"omin katu 2a, FI-00014 Helsinki, Finland\\
  $^{9}$ Leiden Observatory, Leiden University, PO Box 9513, NL-2300 RA Leiden, the Netherlands\\
  $^{10}$ Max Planck Institute for Astrophysics, D-85748 Garching, Germany\\
}
\begin{document}

\maketitle

\begin{abstract}
  \noindent 
  The scaling of disk galaxy rotation velocity with baryonic mass (the
  ``Baryonic Tully-Fisher'' relation; BTF) has long confounded galaxy
  formation models. It is steeper than the $M\propto V^3$ scaling
  relating halo virial masses and circular velocities and its zero
  point implies that galaxies comprise a very small fraction of
  available baryons.  Such low galaxy formation efficiencies may in
  principle be explained by winds driven by evolving stars, but the
  tightness of the BTF relation argues against the substantial scatter
  expected from such vigorous feedback mechanism.  We use the
  APOSTLE/EAGLE simulations to show that the BTF relation is well
  reproduced in $\Lambda$CDM simulations that match the size and
  number of galaxies as a function of stellar mass. In such models,
  galaxy rotation velocities are proportional to halo virial velocity
  and the steep velocity-mass dependence results from the decline in
  galaxy formation efficiency with decreasing halo mass needed to
  reconcile the CDM halo mass function with the galaxy luminosity
  function. Despite the strong feedback, the scatter in the simulated
  BTF is smaller than observed, even when considering all simulated
  galaxies and not just rotationally-supported ones. The simulations
  predict that the BTF should become increasingly steep at the faint
  end, although the velocity scatter at fixed mass should remain
  small. Observed galaxies with rotation speeds below $\sim 40$ km
  s$^{-1}$ seem to deviate from this prediction.  We discuss
  observational biases and modeling uncertainties that may help to
  explain this disagreement in the context of $\Lambda$CDM models of
  dwarf galaxy formation.
\end{abstract}


\section{Introduction}
\label{SecIntro}

The empirical relation between the rotation velocity and luminosity of
disk galaxies is not only a reliable secondary distance indicator
\citep{Tully1977}, but also provides important clues to the total mass
and mass profiles of their host dark matter halos. The Tully-Fisher (TF)
relation has now been extensively studied observationally; its
dependence on photometric passband, in particular, is relatively
well-understood, and the relation is now generally cast in terms of
galaxy stellar mass rather than luminosity \citep[e.g. ][]{McGaugh2000,Bell2001,Pizzagno2005,Torres-Flores2011}.

This relation is well approximated by a single power law with small 
scatter, at least for late-type galaxies with stellar masses $\gtrsim
10^{9.5}\, \rm M_\odot$ and velocities $\gtrsim 65$ ${\rm km}\,{\rm s^{-1}}$ \citep{McGaugh2000,Bell2001}. At
lower masses/velocities, the relation deviates from a simple power
law, presumably because the contribution of cold gas becomes more and
more prevalent in dwarf galaxies. Indeed, the power-law scaling may be
largely rectified at the faint end by considering baryonic
masses rather than stellar mass alone. The ``baryonic Tully-Fisher''
relation, as this relation has become known (or BTF, for short), is
well approximated by a single power law over roughly three decades in
mass and a factor of six in velocity
\citep{McGaugh2000,Verheijen2001,Stark2009}. Its scatter is quite
small, at least when only galaxies with high-quality data and
radially-extended rotation curves are retained for analysis
\citep[][]{McGaugh2012,Lelli2016}. 

The interpretation of the Tully-Fisher relation in cosmologically-motivated models
of galaxy formation has long been problematic. From a cosmological
viewpoint, the Tully-Fisher relation is understood as reflecting the
equivalence between halo mass and circular velocity imposed by the
finite age of the Universe \citep[see,
e.g.,][]{Mo1998,Steinmetz1999}. That characteristic timescale
translates into a fixed density contrast that implies a linear scaling
between virial\footnote{We define the virial mass, $M_{200}$, as that
  enclosed by a sphere of mean density $200$ times the critical
  density of the Universe, $\rho_{\rm crit}=3H^2/8\pi G$.  Virial
  quantities are defined at that radius, and are identified by a
  ``200'' subscript.} radius and velocity, or a simple $M\propto V^3$
relation between mass and circular velocity. A power-law scaling
between galaxy mass and disk rotation velocity is therefore expected
if galaxy mass and rotation speed scale with virial mass and virial
velocity, respectively.

The latter conditions are not trivial to satisfy, as a simple example
illustrates. The Milky Way's baryonic mass is roughly $\approx 6 \times
10^{10}\, M_\odot$ \citep{RixBovy2013}, and its rotation velocity is
approximately constant at $\approx 220$ km/s over the whole Galactic
disk, out to at least $10$ kpc. A halo of similar virial velocity, on
the other hand, has a virial radius of $\approx 310$ kpc and a virial
mass of order $M_{200}\sim 3.5\times 10^{12}\, M_\odot$, or $\approx
6\times 10^{11}\, M_\odot$ in baryons, assuming a cosmic baryon
fraction of $f_{\rm bar}=\Omega_b/\Omega_M=0.17$. A majority of these
baryons can in principle cool and collapse into the Milky Way disk
\citep[see, e.g.,][]{WhiteFrenk1991}. This example illustrates two
important points: (i) only a small fraction of available baryons
are today assembled at the center of the Milky Way halo, and 
(ii) the radius where disk
rotational velocities are measured is much smaller than the virial
radius of its surrounding halo, where its virial velocity is measured.

These points are quite important for models that try to account for
the observed BTF relation. So few baryons assemble into galaxies that
it is unclear how, or whether, their masses should scale with virial
mass. Furthermore, the disk encompasses such a small fraction of the
halo dark matter, and its kinematics probes the potential so far from
the virial boundary, that a simple scaling between galaxy rotation
speed and virial circular velocity might be justifiably
discounted. Finally, it is quite conceivable that the mechanism that
so effectively limits the fraction of baryons that settle into a
galaxy (mainly feedback from evolving stars and supermassive black
holes, in current models) might also exhibit large halo-to-halo
variations due to the episodic nature of the star formation
activity. This makes the rather tight scatter of the observed
Tully-Fisher relation quite difficult to explain \citep{McGaugh2012}.

These difficulties explain why the literature is littered with failed
attempts to reproduce the Tully-Fisher relation in a cold dark
matter-dominated universe. Direct galaxy formation simulations, for
example, have for many years consistently produced galaxies so massive and compact
that their rotation curves were steeply declining and, generally, a
poor match to observation \citep[see, e.g.,][and references
therein]{Navarro2000,Abadi2003,Governato2004,Scannapieco2012}.  Even semi-analytic models,
where galaxy masses and sizes can be adjusted to match observation,
have had difficulty reproducing the Tully-Fisher relation \citep[see,
e.g.,][]{Cole2000}, typically predicting velocities at given mass that
are significantly higher than observed unless somewhat arbitrary
adjustments are made to the response of the dark halo
\citep[][]{Dutton2009}.

The situation, however, has now started to change, notably as a result
of improved recipes for the subgrid treatment of star formation and
its associated feedback in direct simulations. As a result, recent
simulations have shown that rotationally-supported disks with
realistic surface density profiles and relatively flat rotation curves
can actually form in cold dark matter halos when feedback is strong
enough to effectively regulate ongoing star formation by limiting
excessive gas accretion and removing low-angular momentum gas
\citep[see, e.g.,][]{Guedes2011,Brook2012,McCarthy2012,Aumer2013,Marinacci2014}.

These results are encouraging but the number of individual systems
simulated so far is small, and it is unclear whether the same codes
would produce a realistic galaxy stellar mass function or reproduce
the scatter of the Tully-Fisher relation when applied to a
cosmologically significant volume. The role of the dark halo response
to the assembly of the galaxy has remained particularly contentious,
with some authors arguing that substantial modification to the
innermost structure of the dark halo, in the form of a
constant-density core or cusp expansion, is needed to explain the
disk galaxy scaling relations \citep{Dutton2009,Chan2015}, while other
authors find no compelling need for such adjustment \citep[see,
e.g.,][]{Vogelsberger2014,Schaller2015a,Lacey2015}.

The recent completion of ambitious simulation programmes such as the
{\small EAGLE} project \citep{Schaye2015,Crain2015}, which follow the
formation of thousands of galaxies in cosmological boxes $\approx 100$
Mpc on a side, allow for a reassessment of the situation. The subgrid
physics modules of the {\small EAGLE} code have been calibrated to
match the observed galaxy stellar mass function and the sizes of
galaxies at $z=0$, but no attempt has been made to match the BTF
relation, which is therefore a true corollary of the model. The same
is true of other relations, such as color bi-modality, morphological
diversity, or the stellar-mass Tully-Fisher relation of bright
galaxies, which are successfully reproduced in the model
\citep{Schaye2015,Trayford2015}. Combining {\small EAGLE} with multiple
realizations of smaller volumes chosen to resemble the surroundings of
the Local Group of Galaxies \citep[the {\small APOSTLE} project, see,
e.g.,][]{Fattahi2016,Sawala2015b}, we are able to study the resulting
BTF relation over four decades in galaxy mass. In particular, we are
able to examine the simulation predictions for some of the faintest
dwarfs, where recent data have highlighted potential deviations from a
power-law BTF and/or increased scatter in the relation
\citep{Geha2006,Trachternach2009}.

We begin with a brief description of {\small EAGLE} and {\small
  APOSTLE} in Sec.~\ref{SecNumSims} and present our main results in
Sec.~\ref{SecResults}. We investigate numerical
convergence in Sec.~\ref{SecSimGxConv}. The gas/stellar content and
size as a function of galaxy mass are presented in
Sec.~\ref{SecSimGxProp} before comparing the simulated BTF relation
with observation in Sec.~\ref{SecBTF}. We examine the predicted
faint-end of the relation in Sec.~\ref{SecFaintEnd} before concluding
with a brief summary of our main conclusions in Sec.~\ref{SecConc}.
%
\begin{center} \begin{figure*} 
\includegraphics[width=0.49\linewidth]{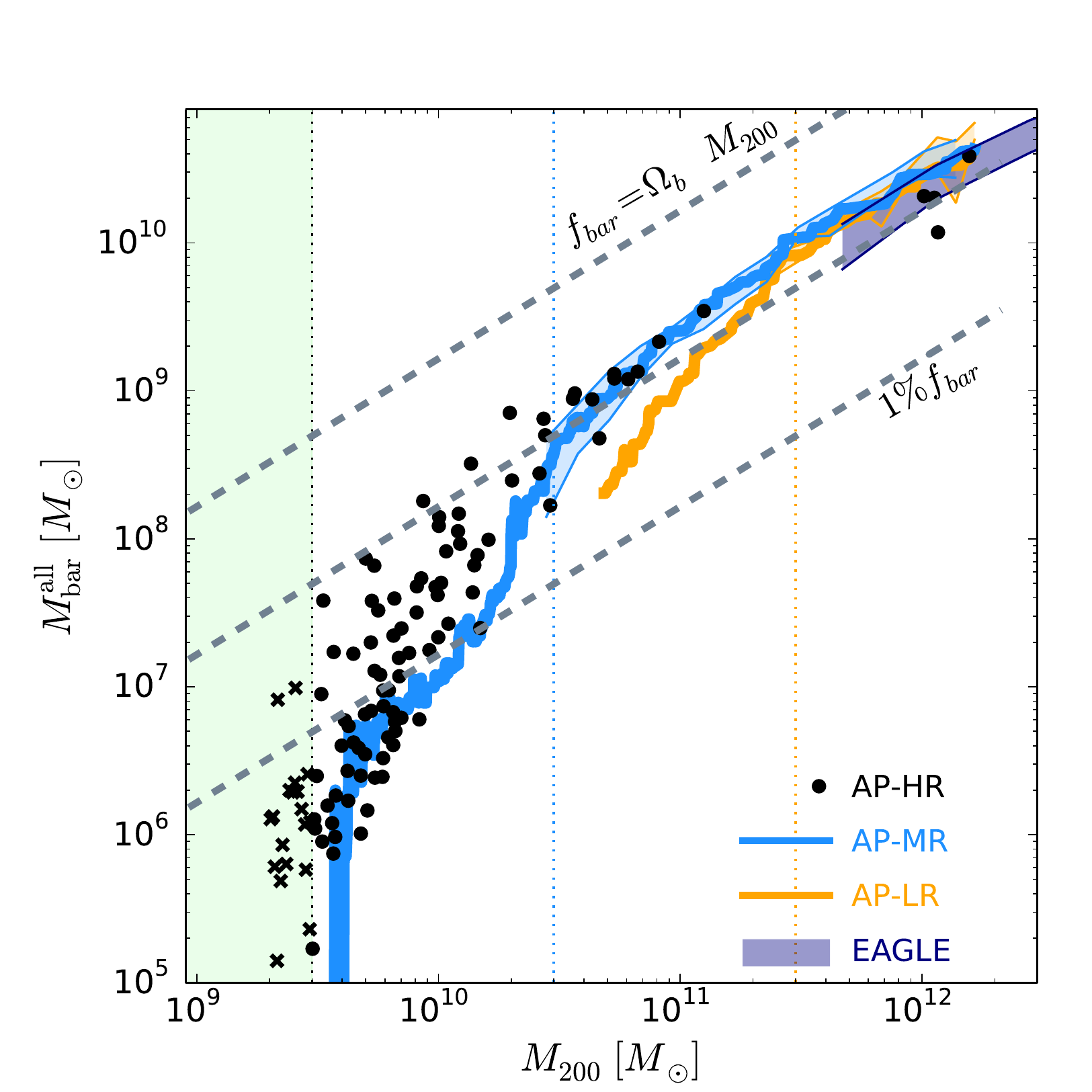}
\includegraphics[width=0.49\linewidth]{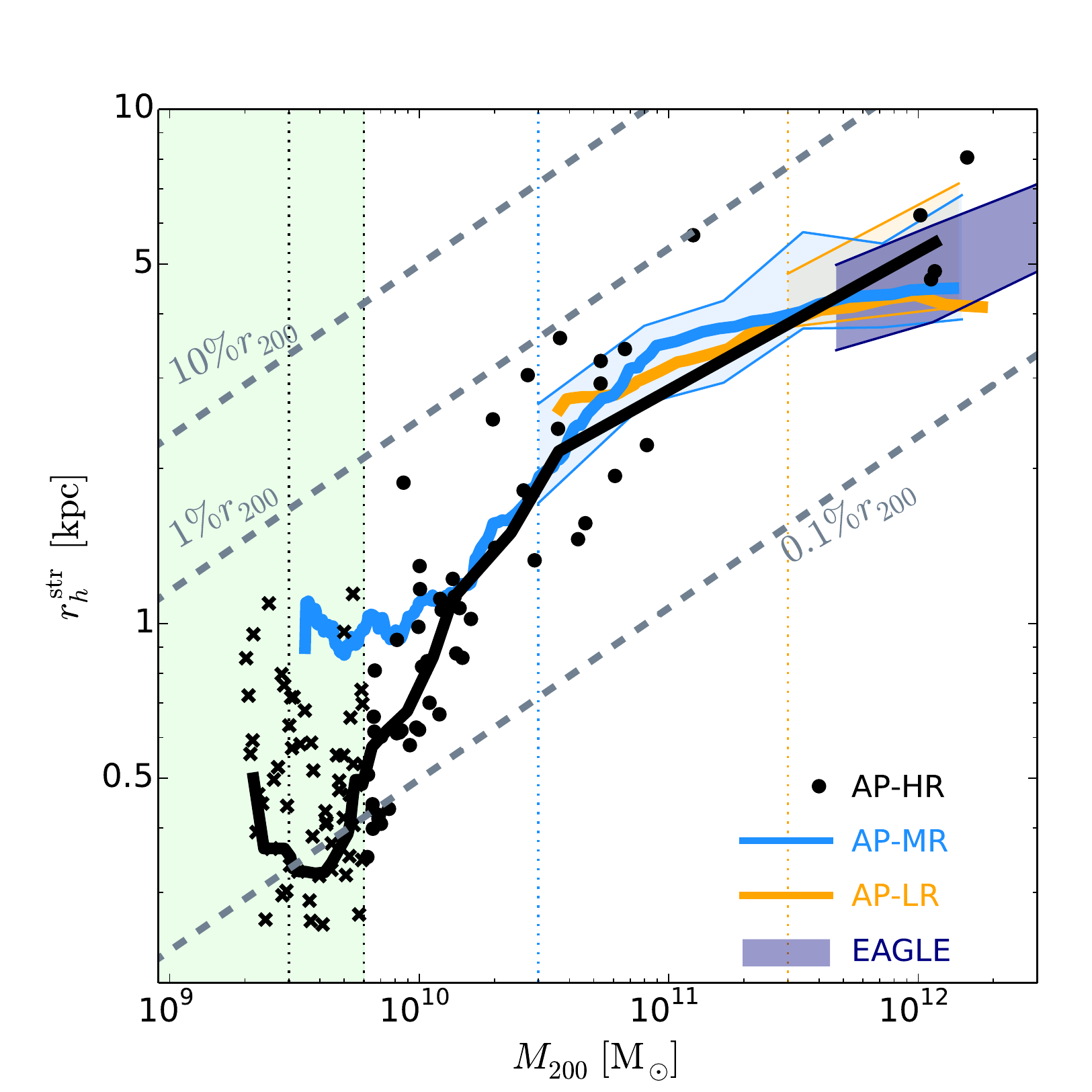}
\caption{{\it Left:} Galaxy baryonic mass
  ($M_{\rm bar}^{\rm all}=M_{\rm gas}^{\rm all} + M_{\rm str}$) vs
  virial mass ($M_{200}$) in our simulated galaxy sample. Shaded
  regions indicate the interquartile baryonic mass range at given
  $M_{200}$ and highlight the virial mass range over which the
  simulation results are insensitive of resolution. Vertical dotted
  lines indicate the minimum converged virial mass for each resolution
  level.  Thick lines of matching color indicate the median trend for
  each simulation set, as specified in the legend, and extend to
  virial masses below the minimum needed for convergence. Dashed grey
  lines indicate various fractions of all baryons within the virial
  radius. Note the steep decline in ``galaxy formation efficiency''
  with decreasing virial mass.  Dark filled circles indicate the
  results of individual AP-HR galaxies. A light green shaded region
  highlights non-converged systems in our highest resolution
  runs.  Crosses are used to indicate galaxies in halos considered
  ``not converged'' numerically.  {\it Right:} Stellar half-mass
  radius, $r_{\rm h}^{\rm str}$, as a function of virial mass for
  simulated galaxies. Symbols, shading and color coding are as in the left
  panel. Limited resolution sets a minimum size for galaxies in poorly
  resolved halos. The same minimum mass needed to ensure convergence in baryonic mass
  seems enough to ensure convergence in galaxy size, except perhaps
    for AP-HR, for which we adopt a minimum
    converged virial mass of $6 \times 10^{9}\, M_\odot$. The
  values adopted for the minimum virial mass are listed in
  Table~\ref{TabSimPar}.}
\label{FigMbarMhalo}
\end{figure*}
\end{center}

%
\begin{center} \begin{figure} 
\includegraphics[width=\linewidth]{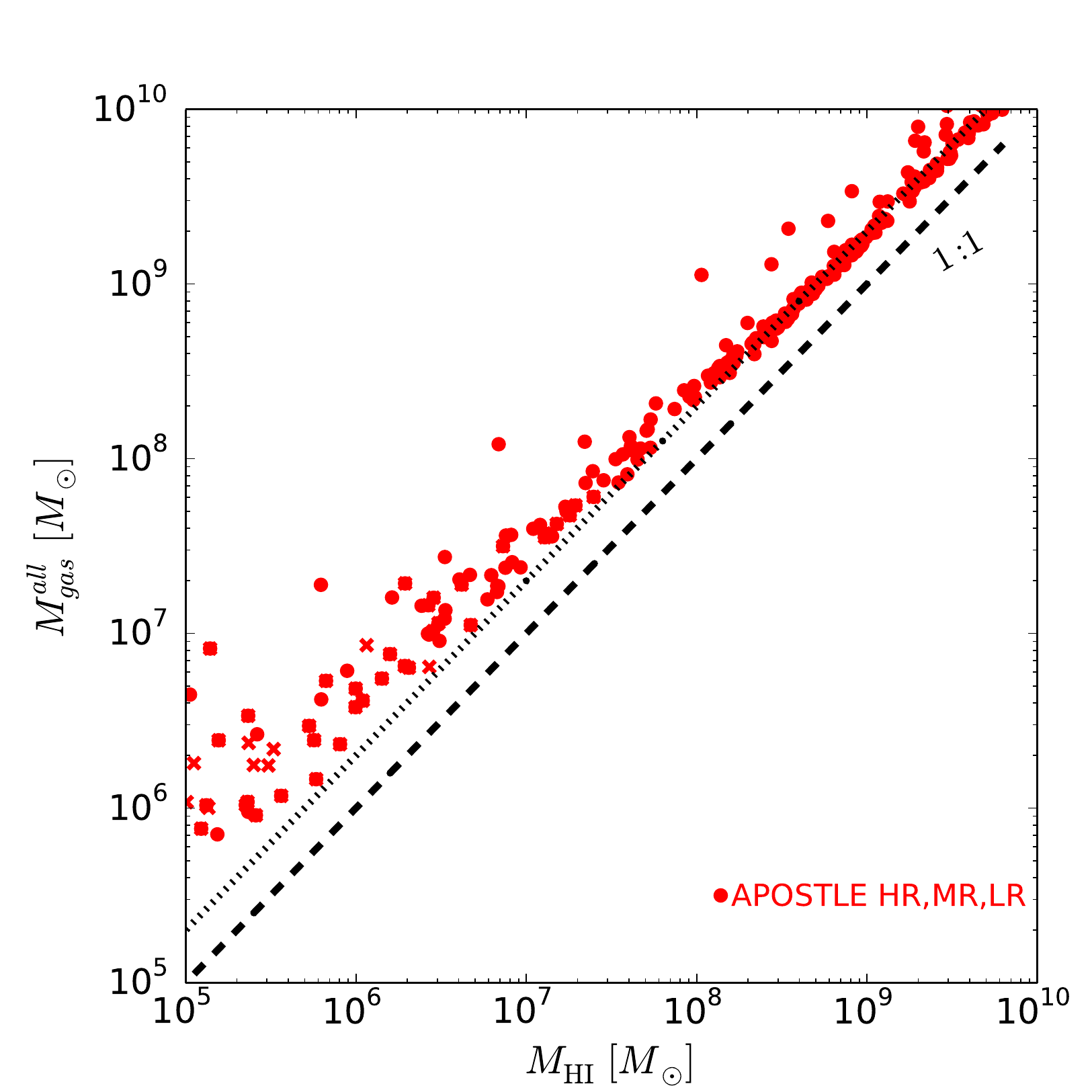}
\caption{Relation between total gas mass $M_{\rm gas}^{\rm all}$ 
and the mass in neutral hydrogen $M_{\rm HI}$ within $r_{\rm gal}$, computed following
the prescription in \citet{Rahmati2013}. The dashed line shows the one-to-one relation
and the dotted line corresponds to $M_{\rm gas}^{\rm all} = 2 \, M_{\rm HI}$,
which is a good approximation for simulated massive galaxies. Low-mass
dwarfs, on the other hand, have a higher fraction of ionized
gas and hence their total gas masses may be substantially
underestimated by applying to them the same HI scaling factor as for
massive systems. As before, crosses are used to indicate galaxies
in the highest resolution runs that are in halos considered 
“not converged” numerically.}
\label{FigMHI}
\end{figure}
\end{center}

\section{Numerical Simulations}
\label{SecNumSims}

\subsection{The Code}

The simulations we use here were run using a modified version of the
SPH code {\sc P-Gadget 3} \citep{Springel2005b}, as developed for the
{\small EAGLE} simulation project
\citep{Schaye2015,Crain2015,Schaller2015c}. We refer the reader to the
main {\small EAGLE} papers for further details, but list here the main
code features, for completeness. In brief, the code includes the
``Anarchy'' version of SPH (Dalla Vecchia, in preparation, see also
Appendix A in \citet{Schaye2015}), which
includes the pressure-entropy variant proposed by \citet{Hopkins2013};
metal-dependent radiative cooling/heating \citep{Wiersma2009a}, reionization
of Hydrogen and Helium (at redshift $z=11.5$ and $z=3.5$,
respectively), star formation with a metallicity dependent density
threshold \citep{Schaye2004,DallaVecchia2008}, stellar evolution and
metal production \citep{Wiersma2009b}, stellar feedback via stochastic
thermal energy injection \citep{DallaVecchia2012}, and the growth of,
and feedback from, supermassive black holes
\citep{Springel2005c,Booth2009,Rosas-Guevara2013}. The free parameters of the subgrid
treatment of these mechanisms in the {\small EAGLE} code have been
adjusted so as to provide a good match to the galaxy stellar mass
function, the typical sizes of disk galaxies, and the stellar
mass-black hole mass relation, all at $z \approx 0$.

\subsection{The Simulations}

We use two sets of simulations for the analysis we present here. One
is the highest-resolution $100$ Mpc-box {\small EAGLE} run
(Ref-L100N1504). This simulation has a cube side length of $100$ comoving
Mpc; $1504^3$ dark matter particles each of mass $9.7 \times 10^{6} M_\odot$ ; the same
number of gas particles each of initial mass $1.8 \times 10^6 M_\odot$; and a
Plummer-equivalent gravitational softening length of $700$ proper pc
(switching to comoving for redshifts higher than $z=2.8$). The
cosmology adopted is that of \citet{PlanckCollaboration2014}, with
$\Omega_M= 0.307$, $\Omega_\Lambda = 0.693$, $\Omega_b = 0.04825$, $h
= 0.6777$ and $\sigma_8 = 0.8288$.

The second set of simulations is the {\sc APOSTLE} suite of zoom-in
simulations, which evolve $12$ volumes tailored to match the spatial
distribution and kinematics of galaxies in the Local Group
\citep{Fattahi2016}. Each volume was chosen to contain a pair of halos
with individual virial mass in the range $5\times 10^{11}$-$2.5 \times
10^{12} \, M_\odot$. The pairs are separated by a distance comparable
to that between the Milky Way (MW) and Andromeda (M31) galaxies ($800
\pm 200$ kpc) and approach with radial velocity consistent with that of
the MW-M31 pair ($0$-$250$ km/s).

The {\small APOSTLE} volumes were selected from the {\small DOVE}
N-body simulation, which evolved a cosmological volume of $100$ Mpc on a
side in the WMAP-7 cosmology \citep{Komatsu2011}.  The {\small APOSTLE} runs were
performed at three different numerical resolutions; low (AP-LR),
medium (AP-MR) and high (AP-HR), differing by successive factors of
$\approx 10$ in particle mass and $\approx 2$ in gravitational force
resolution. All $12$ volumes have been run at medium and low
resolutions, but only two high-res simulation volumes have been
completed. Table~\ref{TabSimPar} summarizes the main parameters of
these simulations.

We use the {\small SUBFIND} algorithm to identify ``galaxies''; i.e.,
self-bound structures \citep{Springel2001,Dolag2009} in a catalog of
friends-of-friends (FoF) halos \citep{Davis1985} built with a linking length
of $0.2$ times the mean interparticle separation. We retain for
analysis only the {\it central} galaxy of each FoF halo, and remove
from the analysis any system contaminated by lower-resolution
particles in the {\small APOSTLE} runs. Baryonic galaxy masses
(stellar plus gas) are computed within a fiducial ``galaxy radius'',
defined as $r_{\rm gal}=0.15\, r_{200}$. We have verified that this is
a large enough radius to include the great majority of the
star-forming cold gas and stars bound to each central galaxy.

\begin{table}
\centering 
\begin{tabular}{l|c| c| r| r} 
\hline
\hline 
& \multicolumn{2}{c}{Particle Masses} & Max Softening & $M_{\rm 200}^{\rm conv}$\\
Label &  DM $[M_\odot]$ & Gas $[M_\odot]$ & [proper pc] & [$M_\odot$] \\  [0.5ex] 
\hline 
AP-HR &  $5.0\times 10^{4}$ &  $1.0\times 10^{4}$ & $94$ & $6.0\times 10^9$ \\ 
AP-MR &  $5.9\times 10^{5}$ &  $1.3\times 10^{5}$ & $216$ & $3.0\times 10^{10}$ \\ 
AP-LR &  $7.3\times 10^{6}$ &  $1.5\times 10^{6}$ & $500$ & $3.0\times 10^{11}$ \\ 
{\small EAGLE} &  $9.7\times 10^{6}$ &  $1.8\times 10^{6}$ & $700$ & $3.0\times 10^{11}$ \\ 
\hline 
\vspace{-.3cm}
\end{tabular}
\caption{Numerical parameters of the {\small APOSTLE} and {\small EAGLE} simulations. 
{\small APOSTLE} simulations
are labeled ``AP" followed by the level of resolution: LR, MR and HR (low, medium and high resolution). The last column summarizes the minimum virial mass required for 
convergence $M_{200}^{\rm conv}$ in each resolution. See Sec. 3.1.} 
\label{TabSimPar} 
\end{table}

\section{Results} 
\label{SecResults}

\subsection{Galaxy formation efficiency and numerical convergence}
\label{SecSimGxConv}

The left panel of Fig.~\ref{FigMbarMhalo} shows the relation between
virial mass, $M_{200}$, and galaxy baryonic mass, $M_{\rm bar}^{\rm all}$, in
our simulations, where $M_{\rm bar}^{\rm all}$ is computed by counting
{\it all} gas and stellar particles within $r_{\rm gal}$. Shaded regions bracket the interquartile range in
$M_{\rm bar}^{\rm all}$ as a function of virial mass for each of the simulation
sets, as indicated in the legend. Thick solid lines of matching colors
indicate the median trend. Individual symbols indicate results for the
high-resolution AP-HR run, since the total number of galaxies in those
two completed volumes is small. 

The dashed grey lines indicate the location of galaxies whose masses
make up $100\%$ (top), $10\%$ (middle) and $1\%$ of all baryons within
the virial radius. Fig.~\ref{FigMbarMhalo} shows that the galaxy
formation efficiency is low in all halos (less than $\approx 20\%$) 
and that it decreases steadily with decreasing virial mass. 
Galaxies in the most massive
halos shown have been able to assemble roughly $15$-$20\%$ of their
baryons in the central galaxy, but the fraction drops to about $1\%$
in $\approx 10^{10}\, M_\odot$ halos for the case of AP-HR. Such a steep
decline is expected in any model that attempts to reconcile the
shallow faint-end of the galaxy stellar mass function with the steep
low-mass end of the halo mass function (see, e.g., the discussion in Sec.~5.2
of \citealt{Schaye2015} and in Sec.~$4$ of \citealt{McCarthy2012}).

The left panel of Fig.~\ref{FigMbarMhalo} also shows the limitations
introduced by numerical resolution. The results for the various
simulations agree for well-resolved halos, but the mean galaxy mass
starts to deviate in halos resolved with small numbers of
particles. This is most clearly appreciated when comparing the results
of the median $M_{\rm bar}^{\rm all}$ for each {\small APOSTLE} simulation
set. AP-LR results, for example, ``peel off'' below the trend obtained
in higher resolution runs for virial masses less than $\approx 3 \times
10^{11}\, M_\odot$. Those from AP-MR, in turn, deviate from the AP-HR
trend below $3 \times 10^{10}\, M_\odot$. To define convergence for
the high resolution run, we simply adopt a similar factor of $10$ between
AP-MR and AP-HR. These limits are shown with
thin vertical lines in the left panel of Fig.~\ref{FigMbarMhalo}.

The issue of convergence in baryonic simulations is complex, since
increasing resolution means that new physical processes enter into
play and it is unclear whether a recalibration of the sub-grid physics
should or should not be performed \citep[see detailed discussion
in][]{Schaye2015}.  We adopt here the simple approach 
of selecting objects for which different resolutions give
consistent baryonic masses. Noting that the particle mass (gas plus
dark matter) is $8.8 \times 10^6\; M_\odot$ and
$6.0 \times 10^4\; M_\odot$ for AP-LR and AP-HR, respectively, a
simple rule of thumb is then that, on average, only halos resolved
with at least $50,000$ particles give consistent galaxy masses for all
runs. We highlight this mass range for each of these runs by shading
the interquartile range above the minimum ``converged'' mass.

The right-hand panel of Fig.~\ref{FigMbarMhalo} examines convergence
in the size of the galaxy (the 3D stellar half-mass radius,
$r_h^{\rm str})$, as a function of virial mass. This panel shows that
galaxy sizes approach a constant value below a certain,
resolution-dependent, virial mass. This may be traced to the combined
effects of limited resolution and of the choice of a polytropic
equation of state for dense, cold gas in the simulations. As discussed
by \citet{Crain2015}, the equation of state imposes an effective
minimum size for the cold gas in a galaxy which explains the constant
size of low-mass galaxies seen in the right-hand panel of
Fig.~\ref{FigMbarMhalo}.  

Notably, the same criterion that ensures convergence for galaxy masses
appears to ensure convergence in size, as shown by the shaded regions
in the right-hand panel, which extend down to the same minimum mass as
in the left-hand panel.  The only exception seems to be AP-HR, where
the minimum size is reached at
$M_{200} \approx 6 \times 10^{9}\, M_\odot$.  We shall hereafter adopt that mass
as the minimum halo mass required for convergence for AP-HR.  We
summarize in Table~\ref{TabSimPar} the minimum virial mass,
$M_{\rm 200}^{\rm conv}$, of simulated galaxies retained for further analysis.

\begin{center} \begin{figure*} 
\includegraphics[width=0.49\linewidth]{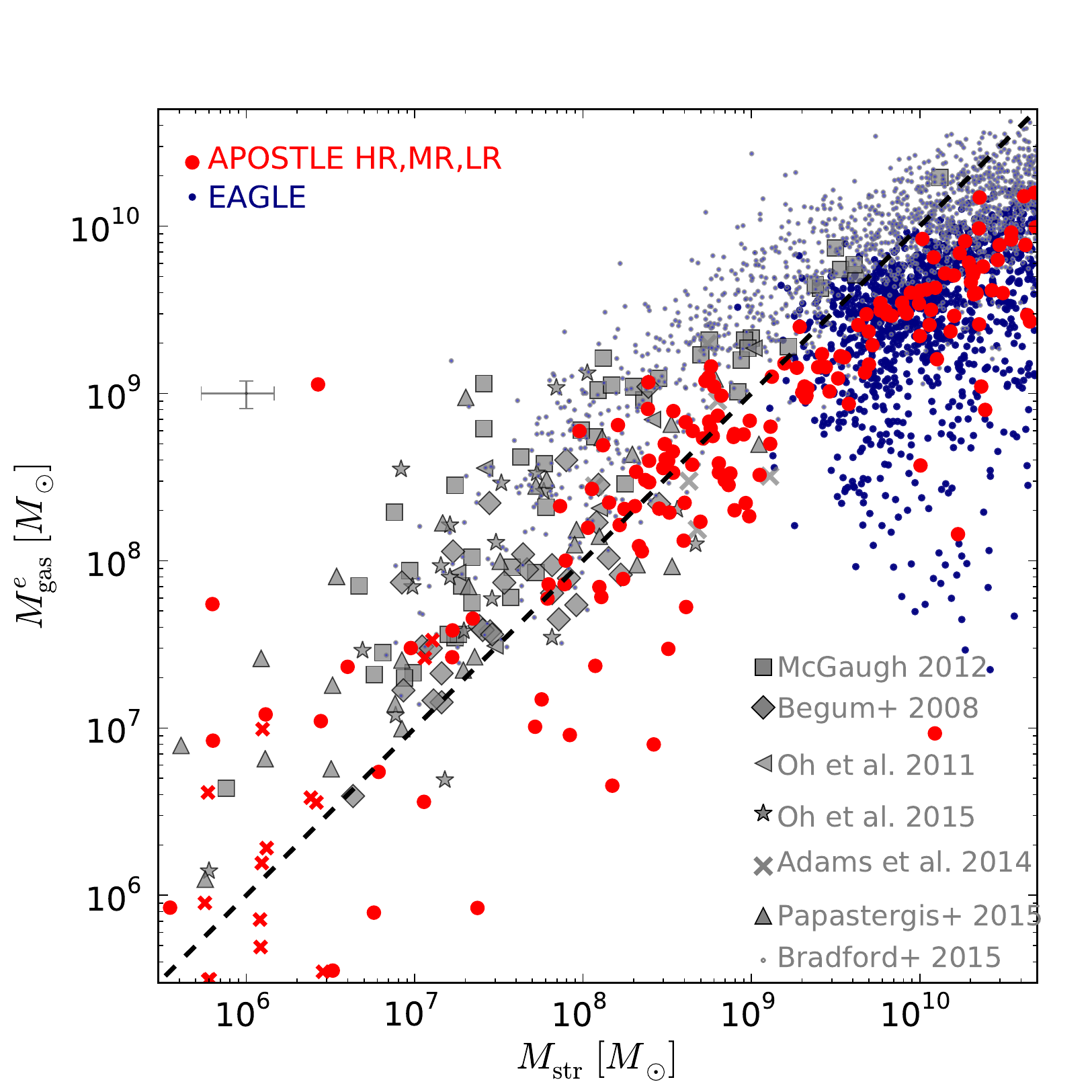}
\includegraphics[width=0.49\linewidth]{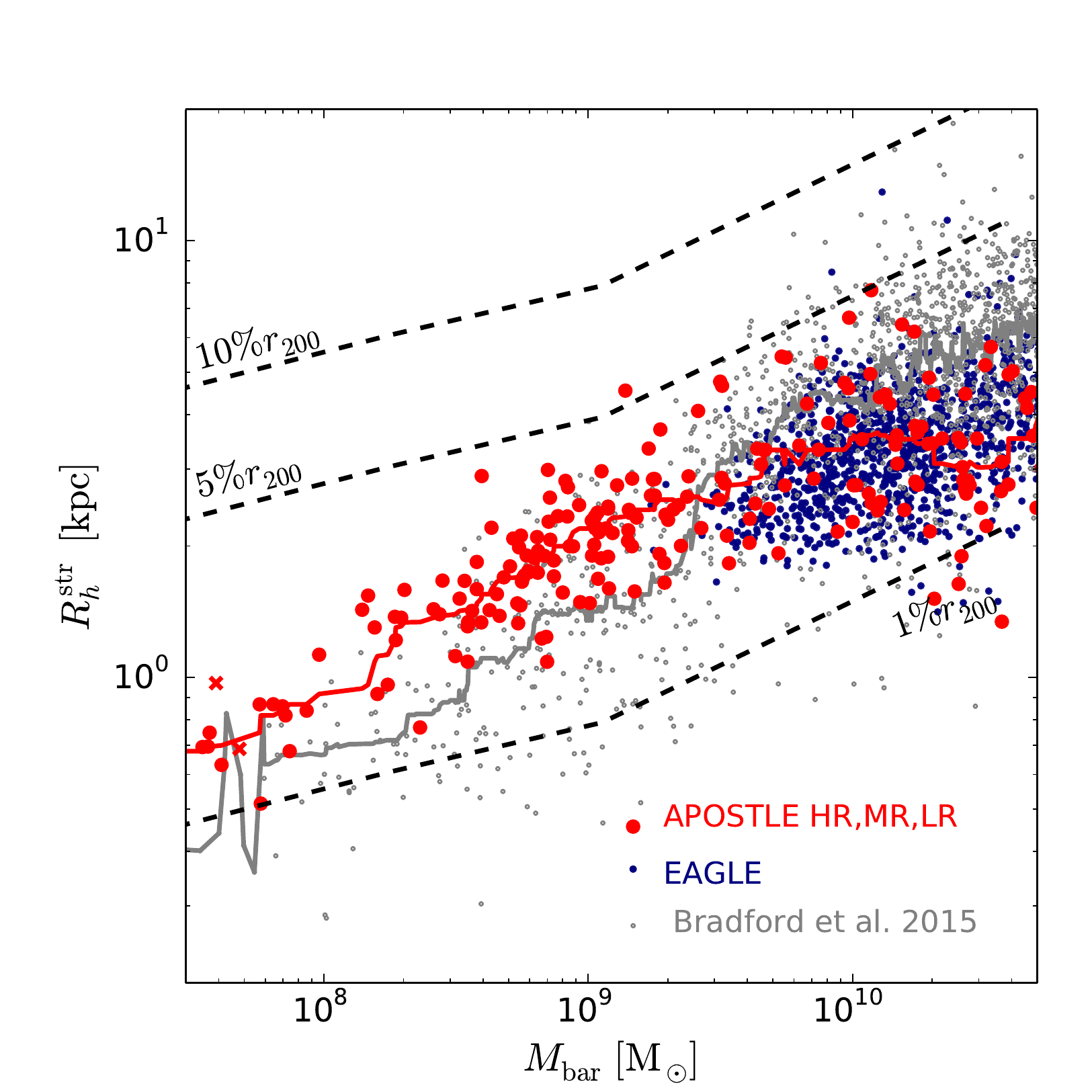}
\caption{Galaxy properties for our sample of simulated galaxies
  (red/blue symbols) and for a compilation of observed galaxies taken
  from the literature (grey symbols). The simulated sample includes
  all galaxies in halos above the corresponding minimum converged
  virial mass (see Table~\ref{TabSimPar}). Crosses are used to denote
  AP-HR galaxies below the minimum converged mass.  {\it Left:}
  Stellar mass vs. gas mass relation, where gas masses in observations
  and simulations are estimated from the HI mass as
  $M_{\rm gas}^e=1.4 \, M_{\rm HI}$. The shape and scatter of the
    relation agree quite well between the simulations and the
    observations, although the simulated ratio of the mass in gas to 
    that in stars is $3$ to $4$ times too small at all stellar masses.
  {\it Right:} Baryonic mass
  ($M_{\rm bar}=M_{\rm str}+M_{\rm gas}^e$) vs projected stellar half-mass
  radius ($R_h^{\rm str}$) relation. Simulated galaxies are compared
  with data from \citet{Bradford2015}. Simulated
    galaxies are somewhat smaller than observed for $M_{\rm bar}$ greater than
    about $2 \times 10^9 \rm M_\odot$, but they are about $50$\% too big at smaller
    baryonic mass.
}
\label{FigMbarMgas}
\end{figure*}
\end{center}
%
\subsection{Gas content and sizes}
\label{SecSimGxProp}

Having established numerical convergence criteria for the baryonic
mass and size of simulated galaxies---two of the most important
ingredients of the BTF relation---we now assess whether ``converged''
galaxies compare favourably with observation in terms of their gas,
size, and stellar content. Estimates of gas mass in observations are
usually derived directly from measurements of neutral hydrogen scaled
up by a factor of $\approx 1.33$ or $1.4$ in order to take into
account the contribution of helium and heavier elements. We note,
however, that such a procedure can seriously underpredict the total
amount of gas, especially for low-mass galaxies, where ionized gas is
expected to be an important contributor. Fig.~\ref{FigMHI} shows the
comparison between the total amount of gas within $r_{\rm gal}$ and
that in neutral hydrogen (HI) for our simulated galaxies ($M_{\rm HI}$
is computed by applying the prescription presented in Appendix A.2 of
\citealt{Rahmati2013}). At high masses the relation is linear with
$M_{\rm gas} \approx 2\, M_{\rm HI}$ (dotted line), but below
$M_{\rm gas}\sim 10^8 \, M_\odot$, a simple scaling of the neutral
hydrogen mass can severely underestimate the total amount of gas in
the galaxy due to increasing importance of ionised gas \citep[see
also][]{Gnedin2012}.

We choose therefore to mimic established practice and, in what
follows, we shall estimate gas masses in simulated galaxies by
$M_{\rm gas}^e = 1.4 M_{\rm HI}$ in order to compare with
observations. We emphasize, however, that none of our conclusions
would change qualitatively if we had used the total amount of gas instead.

Fig.~\ref{FigMbarMgas} shows the gas vs stellar mass (left panel), as
well as the baryonic mass vs (projected) stellar half-mass radius $R_{h}^{\rm str}$ 
of simulated galaxies, compared with a compilation of observational surveys, as
listed in the legends\footnote{Data taken from \citet{Begum2008,
    Oh2011, McGaugh2012, Adams2014, Oh2015,
    Bradford2015}. Additionally, we include a subset of the galaxy
  compilation in \citet{Papastergis2015a}, including data from 
  the ``Survey of HI in Extremely Low-mass Dwarfs, SHIELDS"
  \citep{Cannon2011}, the ``Local Volume HI Survey, LVHIS" 
  \citep{Trachternach2009,Kirby2012}, and Leo P
  \citep{Bernstein-Cooper2014}}. Note that for consistency with observations, 
in the right panel of Fig.~\ref{FigMbarMgas} sizes in the simulations are computed
in 2D, by projecting all star particles along a random line-of-sight direction.

This comparison shows that our galaxies reproduce observed trends
relatively well, although some differences are worth pointing out.
One is the characteristic galaxy mass below which the gas content
dominates the baryonic inventory of a galaxy, which happens for
$M_{\rm str} \lesssim 5 \times 10^{9}\, M_\odot$ in observed galaxies.
In the simulations, although gas is more abundant in low-mass galaxies
than in large ones, it rarely dominates the baryonic mass, with an
average contribution of about half for galaxies with stellar masses
$\lesssim 1 \times 10^{9}\, M_\odot$. 
On average, $M_{\rm gas}^e/M_{\rm str}$ is
a factor $3$-$4$ smaller in simulations than in observations
at fixed $M_{\rm bar}^e$. Interestingly, this factor is independent of
baryonic mass, suggesting that the star-formation efficiency in the 
simulations may be too high by a similar factor at all masses.

The second point to note is that the stellar component of simulated
galaxies at the faint end tends to be slightly larger in size, at fixed
$M_{\rm bar}$, than for the observed counterparts ($\approx 50\%$ effect). 
The reasonable agreement in mass and sizes between observations
and simulations implies that estimates of the disk circular velocity from our
simulations can reliably be compared with observational measurements.

\subsection{The Baryonic Tully-Fisher relation}
\label{SecBTF}

\begin{center} \begin{figure} 
\includegraphics[width=\linewidth]{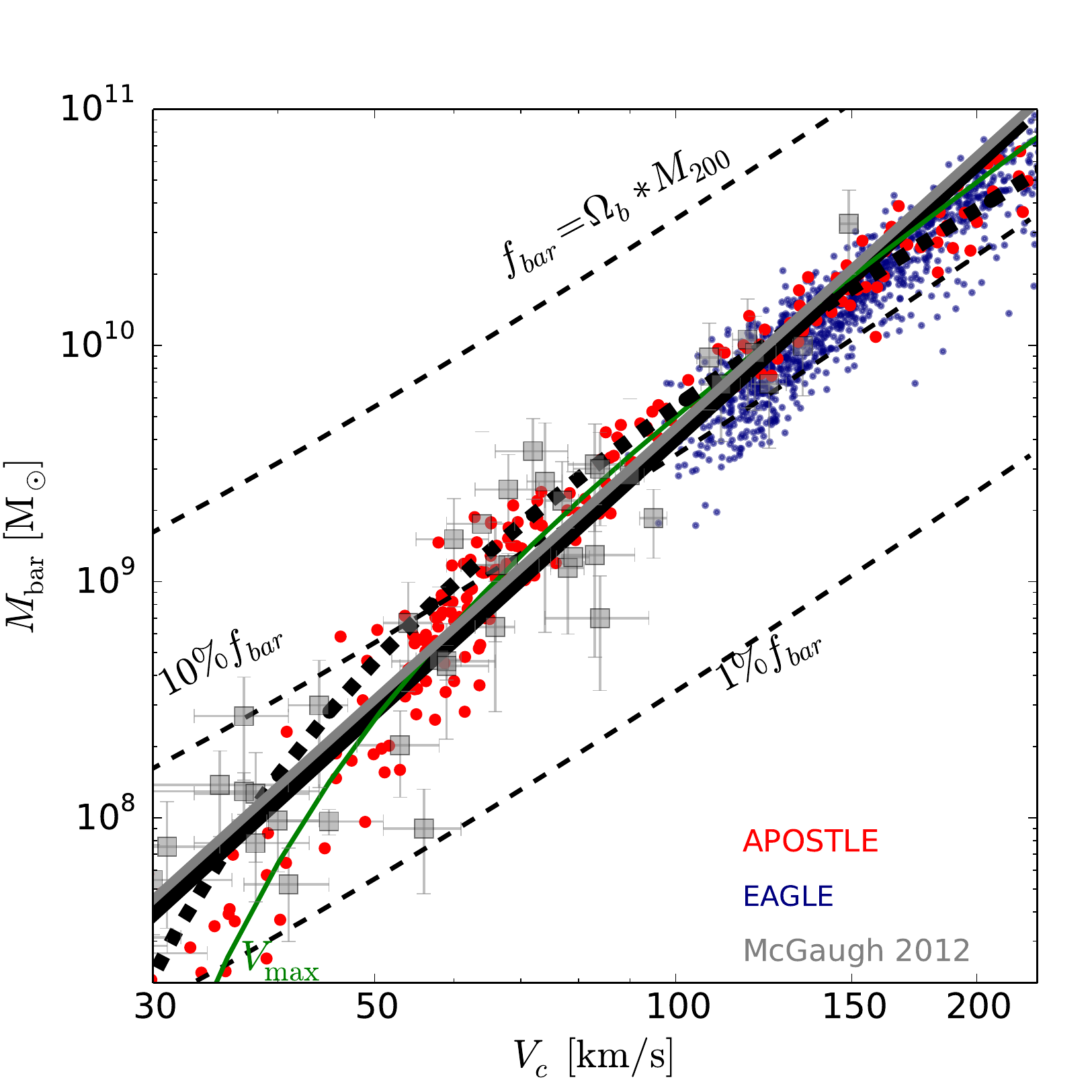}
\caption{Simulated BTF relation for galaxies with circular velocities
  in the range $30<V_{\rm c}/$ km $\rm s^{-1}$/ $<230$. Symbols and colors are as in
  Fig.~\ref{FigMbarMgas}. We use the circular velocity at $2\,
  r_h^{\rm bar}$ for simulated galaxies.  Grey symbols with error bars
  show data from the observational compilation of \citet{McGaugh2012},
  which use the asymptotic ``flat'' rotation velocity.  Simulations
  and observations are generally in good agreement. Solid lines 
   indicate the best power-law fits to simulations (black) 
  and observations (grey). The thick black dashed line is a fit to simulated data
  with steepening slope at the faint end. Parameters of the fit are
  listed in Table~\ref{TabBestFitPar}. The green thin curve depicts the
  relation between baryonic mass and the maximum circular velocity of the 
  halos in our simulations.  Note that the slope of the
  simulated BTF relation is steeper than $V^3$, as a result of the
  declining galaxy formation efficiency in low-mass halos shown in
  Fig.~\ref{FigMbarMhalo}.}
\label{FigBTF}
\end{figure}
\end{center}

We now proceed to examine the velocity scaling of the baryonic
masses. The simulated baryonic Tully-Fisher relation is shown in
Fig.~\ref{FigBTF}, where we plot the circular velocity ($V_{\rm c}^2 = GM(r)/r$) 
estimated at twice the baryonic half-mass radius, $V_{\rm c}(2\, r_{h}^{\rm bar})$, vs
$M_{\rm bar}$, the sum of the stellar and gas mass (computed as
$M_{\rm gas}^e=1.4 M_{\rm HI}$) within $r_{\rm gal}$. We focus
in this figure on galaxies with rotation speeds exceeding $30$
km/s, which includes most galaxies traditionally used in BTF
observational studies, and defer to the next section the analysis of
the relation at the very faint end.

A power law fit to the simulated BTF over this
velocity range suggests a relation $M_{\rm bar}=4.4\times 10^9 (V_{\rm
  c}/100$ km/s$)^{3.6}\, M_\odot$ (as shown by the black solid line). This may
be compared with data for individual galaxies from the compilation of
\citet{McGaugh2012}, which are shown by the grey squares with error
bars, as well as with the power-law fit provided by that reference,
$M_{\rm bar}=4 \times 10^9 (V_{\rm c}/100$ km/s$)^{3.8}\, M_\odot$, 
indicated by the thick grey line.  

The differences between the simulated and observed BTF relations are not
large, especially considering that we are using {\it all} simulated
galaxies in the comparison, without selecting for gas content, size,
morphology, or rotational support.  Galaxies in the observed
compilation, on the other hand, are mainly disk systems where the gas
component dominates\footnote{These were purposefully chosen to be
  gas-dominated in order to minimize uncertainties on their baryonic
  masses that arise from poorly-constrained stellar mass-to-light
  ratios.} and where the rotation curve extends sufficiently far to
reach the asymptotic maximum of the rotation curve. Although we do not
attempt to match such selection procedures in the simulations, the
offset between observed and simulated BTF relations is quite small (at
most $20\%$ in velocity for galaxies with baryonic mass of order
$10^8\, M_\odot$) and would only improve further if we used the
maximum asymptotic velocity for the simulated galaxies. The latter is shown
by the thin green solid line labelled ``$V_{\rm max}$'', which shows 
the fit to the median relation between $M_{\rm bar}$ and $V_{\rm max}$  
as given in Table ~\ref{TabBestFitPar} (see also \citealt{Oman2015}).

%

The agreement between simulated and observed BTF relations
shown in Fig.~\ref{FigBTF} seems to arise naturally in $\Lambda$CDM
simulations that broadly reproduce the galaxy mass function and the
sizes of galaxies as a function of mass. Its normalization at the
luminous end is determined primarily by the need to match the number
density of $L_*$ galaxies. This fixes the average galaxy formation
efficiency in halos of virial mass $\approx 2\times 10^{12}\, M_\odot$,
assigning them a galaxy like the Milky Way (i.e., $M_{\rm bar}\approx
6\times 10^{10}\, M_\odot$). 

\begin{center} \begin{figure} 
\includegraphics[width=\linewidth]{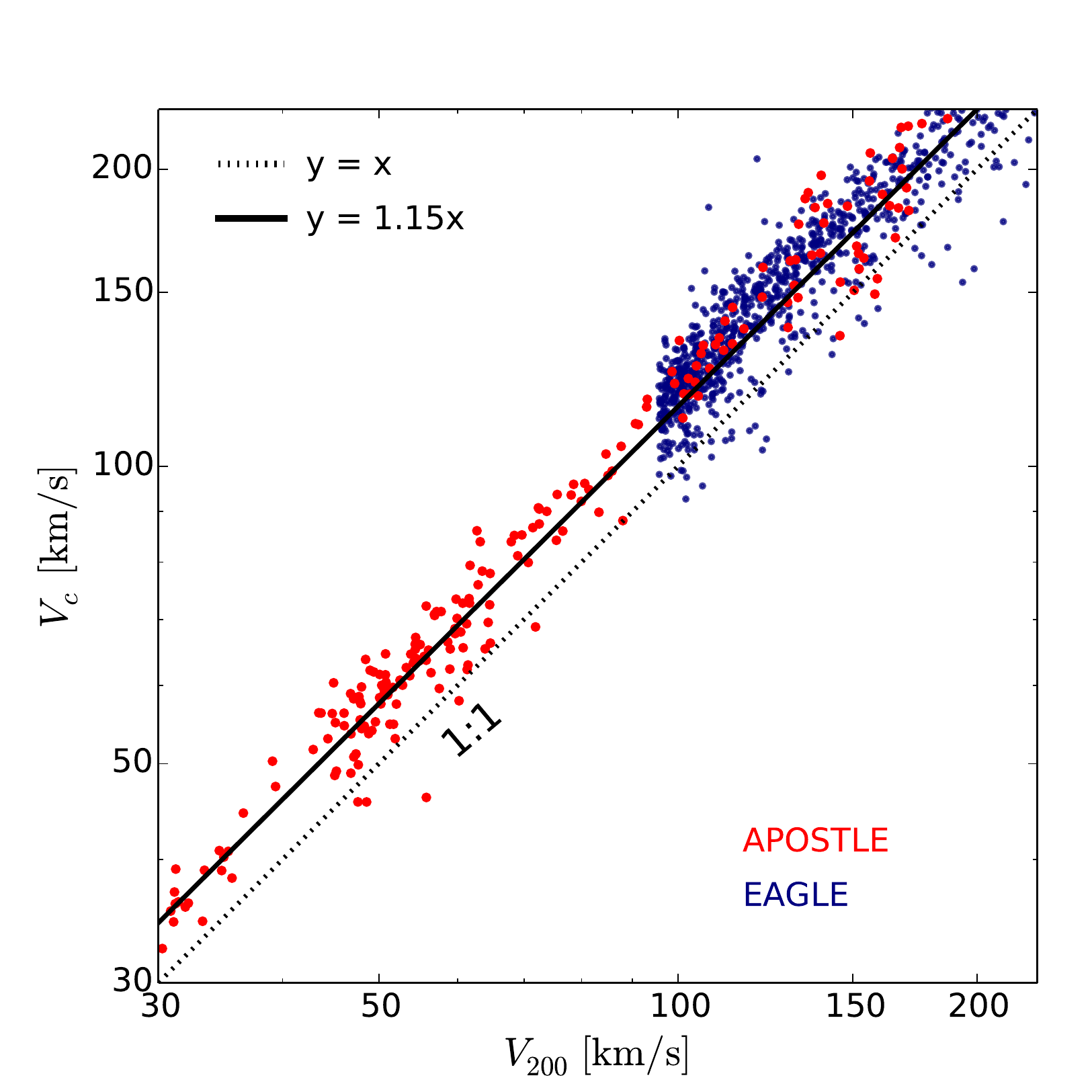}
\caption{Circular velocity at twice the baryonic half-mass radius,
  $V_c(2\, r_h^{\rm bar})$, as a function of virial velocity,
  $V_{200}$, for simulated galaxies. Symbols are as in
  Fig.~\ref{FigMbarMgas}. Note that a simple proportionality
  ($V_c \approx 1.15\, V_{200}$) links these two measures of circular
  velocity (solid black line). 
  This is not a trivial result, but rather a consequence of
  the self-similar nature of $\Lambda$CDM halos, of the smooth decline
  in galaxy size with baryonic mass (Fig.~\ref{FigMbarMgas}), and of the
  mild response of the halo to galaxy assembly. See text for more
  details.}
\label{FigVcV200}
\end{figure}
\end{center}

The virial velocity of such halos, $V_{200}\approx 190 $ km/s, is only
slightly below the $220$ km/s derived from the observed BTF relation
(grey line in Fig.~\ref{FigBTF}), implying that agreement between
simulation and observation follows if the circular velocity
traced by the baryons is approximately $15\%$ greater than their virial velocity.
This is indeed the case for the whole sample, as shown
Fig.~\ref{FigVcV200}, where we plot the circular velocity at
$2\, r_h^{\rm bar}$ as a function of virial velocity for all simulated
galaxies. A simple proportionality links, on average, these two
measures; i.e., $V_{\rm c}(2\, r_h^{\rm bar}) \approx 1.15 \, V_{200}$ (see
thick dashed line in Fig.~\ref{FigVcV200} showing the median of the points 
very close to the one-to-one line), exactly what is needed to
reconcile the normalization of the simulated and observed BTF relations.

We emphasize that this is not a trivial result, but rather a
consequence of the combined effects of (i) the self-similar nature of
$\Lambda$CDM halos, which regulates the total amount of dark matter
enclosed within the luminous region of a galaxy; (ii) the mass and
size of the galaxy, which specifies the baryonic contribution to the
disk rotation velocity; and (iii) the response of the dark halo to the
assembly of the galaxy, which determines how the halo
contracts/expands as baryons collect at its center. The agreement
between observed and simulated BTF shown in Fig.~\ref{FigBTF} should
therefore be considered a major success of this $\Lambda$CDM model of
galaxy formation.

The simulations also clarify why the simulated BTF relation is steeper
than the ``natural'' $M\propto V^3$ relation discussed in
Sec.~\ref{SecIntro}. Since rotation velocities are directly
proportional to virial velocity (Fig.~\ref{FigVcV200}), the steeper
slope mainly reflects the fact that galaxy formation efficiency
declines gradually but steadily with decreasing halo mass, as required
to match the faint-end of the galaxy stellar mass function (see
left-hand panel of Fig.~\ref{FigMbarMhalo}).

The response of the dark halo in the {\small EAGLE/APOSTLE}
simulations has been discussed in detail in \citet{Schaller2015a,Schaller2015b}. We shall not
repeat that analysis here, except to point out that, for radii as large
as $2\, r_h^{\rm bar}$, it can be characterized fairly accurately by
some mild ``adiabatic'' contraction, which is only noticeable in the
most massive, baryon-dominated galaxies. The galaxy formation
efficiency in dwarf galaxy halos is so low that their inner dark mass
profiles are unaffected by the assembly of the galaxies \citep[see
also the discussion in][]{Oman2015}.

Finally, we consider the scatter in the simulated BTF relation. A
conservative estimate may be derived by considering {\it all}
simulated galaxies, regardless of morphological type, size, gas
fraction, or rotation curve shape. We find an rms scatter of $0.20$ dex
in mass and $0.05$ dex in velocity from the best power-law fit to the
data shown in Fig.~\ref{FigBTF}. The scatter is even smaller if,
instead of a power law, one considers a relation whose slope steepens
slightly toward the faint end\footnote{These fits are of the form
  $M_{\rm bar}/{\rm M}_\odot=m_0 \, \nu^{\alpha} \exp(-\nu^{\gamma})$,
  where $\nu$ is the velocity in units of $50$ km/s. The best fit
  parameters $m_0$, $\alpha$, and $\gamma$ are listed in
  Table~\ref{TabBestFitPar}.}. The scatter about this relation is just
$0.14$ dex in mass and $0.04$ dex in velocity. For comparison, the
power-law scatter in the observed BTF relation shown in
Fig.~\ref{FigBTF} is $0.2$ dex in mass and $0.06$ dex in velocity. We
conclude that our simulations have no obvious difficulty accounting
for the small scatter in the BTF relation. Actually, the opposite
seems true once fainter galaxies are considered, as we discuss next.

\begin{table}
  \caption{Best fit parameters to the relation
  $M_{\rm bar}/{\rm M}_\odot=m_0 \, \nu^{\alpha} \exp(-\nu^{\gamma})$,
  where $\nu$ is the velocity in units of $50$ km/s, and can refer to
  either $V_{\rm max}$ or $V_{\rm out}$ as indicated.}
\centering 
\begin{tabular}{l|c|c|c} 
\hline
\hline 
Relation & $m_0$ & $\alpha$ & $\gamma$ \\
 & $[M_\odot]$ &  &  \\
\hline 
$V_{\rm max}$ & $7.1\times 10^8$ & 3.08 & -2.43\\
$V_{\rm out}$ & $1.25 \times 10^9$ & 2.5 & -2.00 \\
\hline 
\vspace{-.3cm}
\end{tabular}
\label{TabBestFitPar} 
\end{table}

\subsection{The faint end of the BTF relation}
\label{SecFaintEnd}

The discussion of the previous section has important consequences for the
faint end of the BTF relation. If galaxy formation efficiencies drop
ever more rapidly with decreasing halo mass (as shown in the left-hand
panel of Fig.~\ref{FigMbarMhalo}), a sharp steepening of the BTF
relation should be expected at the faint end. This is shown explicitly
by the thin green line labeled ``$V_{\rm max}$'' in Fig.~\ref{FigBTF}, which
shows the median baryonic mass as a function of the maximum
asymptotic circular velocity for simulated galaxies. The faint-end
steepening is a direct consequence of the increased efficiency of
feedback in shallower potential wells, and is therefore a robust
prediction of the model.

In order to examine whether this prediction agrees with observation,
we need to extend the observational sample to include fainter galaxies
than those listed in the compilation of \citet{McGaugh2012}. We
therefore add to the observed sample galaxies from the THINGS
\citep{deBlok2008,Oh2011} and LITTLE THINGS \citep{Oh2015} surveys;
those from the compilation of \citet{Papastergis2015a}; as well as
individual galaxies observed by \citet{Begum2008} and
\citet{Adams2014}.

\begin{center} \begin{figure} 
\includegraphics[width=\linewidth]{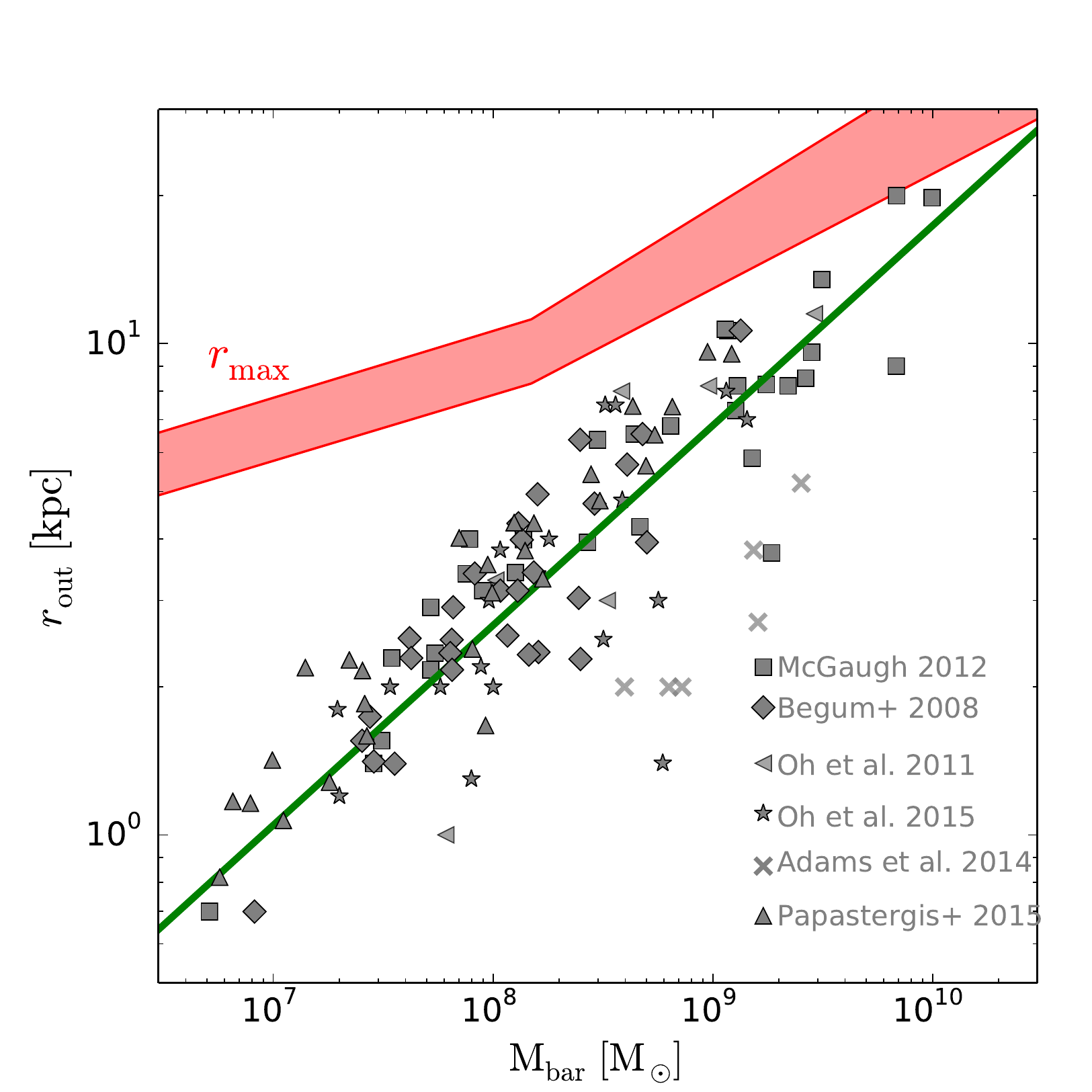}
\caption{The outermost radius, $r_{\rm out}$, of the rotation curve of
  observed galaxies as a function of baryonic mass. The green line shows a power law fit to the
  observations $\rm log(r_{\rm out}/\rm kpc) = 0.42 \, \rm log(M_{\rm bar}/\rm M_\odot) - 2.82$.
  The red-shaded
  area brackets, at given $M_{\rm bar}$, the interquartile range of
  radii where simulated galaxies reach their peak circular
  velocity. The limited extent of observed galaxy rotation curves
  samples only the inner, rising part, of the halo circular velocity
  curve, especially at the low-mass end, 
  $M_{\rm bar}<5 \times 10^8 \rm M_\odot$. This effect needs to be taken
  into account when comparing the faint end of the observed and
  simulated BTF relations. }
\label{FigMbarRout}
\end{figure}
\end{center}

One issue that arises when enlarging the dwarf galaxy sample in this
manner is that in many cases the rotation curve is still rising at the
outermost measured radius and therefore the reported velocity may fall
short of the maximum asymptotic value of the system. One way of
appreciating this is shown in Fig.~\ref{FigMbarRout}, where we plot
the outermost radius of the rotation curve, $r_{\rm out}$, vs baryonic
mass for all the galaxies in the references listed above. Clearly, the
observed radial extent of the rotation curve correlates strongly with
baryonic mass: the lower the galaxy mass the smaller the galaxy and
the shorter its rotation curve. This is in sharp contrast with the
radius, $r_{\rm max}$, at which the maximum circular velocity is
reached in simulated galaxies, which flattens out at small masses as a
result of the steepening of the $M_{\rm bar}$-$M_{200}$
relation. Broadly speaking, all dwarfs in Fig.~\ref{FigMbarRout} with
baryonic masses below $\sim 10^8\, M_\odot$ inhabit halos of similar
virial mass, which results in the very weak dependence of
$r_{\rm max}$ with $M_{\rm bar}$ shown in this figure.

Because $r_{\rm out}$ is in many cases much smaller than
$r_{\rm max}$, especially at the low-mass end, it is important that
velocities are estimated at similar radii when comparing with
observations. We attempt to do this by choosing a value of
  $r_{\rm out}$ for each simulated galaxy based on its baryonic mass
  $M_{\rm bar}$ and by randomly sampling the $M_{\rm
    bar}$-$r_{\rm
    out}$
  relation shown in Fig.~\ref{FigMbarRout}. (In practice we use the
  power-law fit shown by the solid green line and a Gaussian scatter
  in radius of $0.15$ dex.).  This procedure ensures that that
circular velocities are measured for our simulated galaxies at the
same radii on average as for observed galaxies of the same baryonic
mass. Note that for the small masses of most interest in this
  paper, both simulated and observed systems are usually dominated by
  dark matter, so that that the actual distribution of baryons has
  little effect.

\begin{center} \begin{figure} 
\includegraphics[width=90mm]{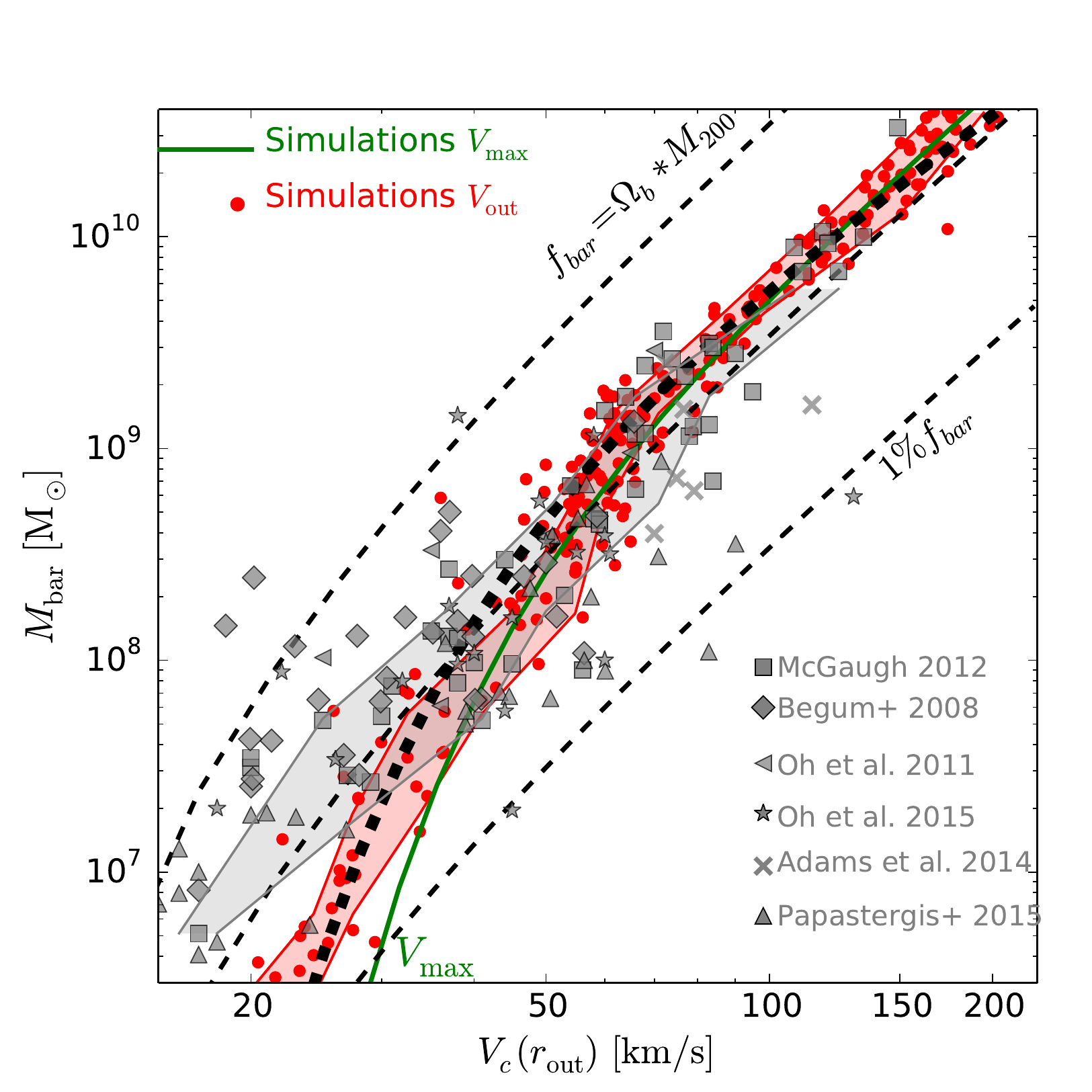}
\caption{Comparison between predicted and observed baryonic
  Tully-Fisher relations, extended to include fainter galaxies than in
  Fig.~\ref{FigBTF}. Grey symbols indicate the observed compilation,
  from references listed in the legend. Velocities are now defined at
  $r_{\rm out}$, the outermost point of the observed rotation curve given in
  Fig.~\ref{FigMbarRout} (or its maximum value, when the two do not
  coincide). Median values of maximum velocity at given baryonic mass 
  for simulated galaxies are
  indicated by the thick solid line labelled ``$V_{\rm max}$''. Small
  dots indicate the predicted velocities of simulated galaxies measured 
  at the $r_{\rm out}$ based on the best power-law fit to the observed sample (see
  green line in Fig.~\ref{FigMbarRout}).
  The shaded areas correspond to the interquartile velocity range at a given
  fixed baryonic mass for the simulated (red) and observed (gray) samples. 
  As expected from Fig.~\ref{FigMbarRout}, $V_c(r_{\rm out})$
  underestimates the maximum velocity in low-mass galaxies by a factor
  of $\approx 1.5$. 
  Note as
  well that the simulated BTF shows a clear steepening in mass at the faint end
  which is less pronounced in the observed BTF. The observed BTF also
  has substantially larger scatter at the faint end, with a number of
  clear outliers with no counterparts in the simulated sample (see text for
  more details).
} 
\label{FigBTFFE}
\end{figure}
\end{center}

The result of this exercise is shown in Fig.~\ref{FigBTFFE}, where
red solid circles show the predicted $V_{\rm out}$ for
simulated galaxies and the shaded areas bracket the
interquartile velocity distribution obtained (at fixed $M_{\rm bar}$).
The comparison illustrates a couple of interesting
points. One is that, as expected, the rotation velocities at the faint
end underestimate the maximum circular velocities by, at times, a
fairly large factor. This has the effect of largely rectifying the BTF
steepening predicted when using $V_{\rm max}$ so that the relation at
the faint end appears to follow a power-law scaling similar to that of
the more luminous systems \citep[see][for a similar analysis and
conclusion]{Brook2015}. The sharp steepening of the BTF at the faint
end may thus be somewhat ``hidden'' by the fact that the sizes of
galaxies scale strongly with baryonic mass, leading to a
systematically larger underestimation of the maximum circular velocity
with decreasing galaxy mass \citep[see discussion in ][]{Papastergis2015b}.

Fig.~\ref{FigBTFFE} also highlights a few differences between the
observed and simulated the BTF relation at the faint end: (i) the
scatter in velocity at given mass is significantly larger in
observations, and (ii) the steepening trend seen in the simulated BTF
faint end is more pronounced than observed, despite the
rectifying effect caused by the small values of $r_{\rm out}$
discussed above.

The first point may be best appreciated by considering the dispersion
in simulated velocities at given mass, which is basically independent of
$M_{\rm bar}$ and only of order $0.05$ dex when measured from the
best-fit  function shown by the thick black dashed line. On the other
hand, the dispersion in the observed data is much greater: for
example, at $M_{\rm bar} \approx 10^8\, M_\odot$ the velocity rms is $0.13$
dex, with some obvious outliers in the observed sample for which there
are no simulated counterparts.

The second point suggests that, at the very faint end, observed
galaxies inhabit halos of lower mass (or lower circular velocity, to
be more precise) than predicted by the model, a result reminiscent of
previous conclusions \citep{Ferrero2012,Papastergis2015a}.
A similar issue arises
when comparing the predicted and estimated masses of Milky Way
satellites, also known as the ``too-big-to-fail'' problem
\citep{Boylan-Kolchin2011}. The difference is, however, small: the
median velocity predicted for galaxies in the range $5\times 10^6 <
M_{\rm bar}/M_\odot < 3\times 10^7$ is $\approx 22$ km/s while the observed
value is $\approx 19$ km/s. 

The observed outliers in Fig.~\ref{FigBTFFE} present a more worrying
puzzle. The ones to the right of the simulated trend can in principle
be explained as baryon-dominated galaxies where the central
concentration of baryonic matter leads to rotation velocities that
exceed the asymptotic maximum of the halo (i.e., ``declining''
rotation curves). The ones on the left, on the other hand, are more
difficult to explain. These are galaxies that are extremely massive in
baryons for their rotation speed or, alternatively, that rotate much
more slowly than is typical for their mass.

Because galaxy formation efficiencies are, almost without exception,
very low in simulated dwarf galaxies, the observed baryonic mass of a galaxy
places a strong lower limit on the virial mass of the halo it
inhabits. The mass of the halo then constrains the total amount of
dark matter within $r_{\rm out}$, placing a strong lower limit on the
circular velocity there. The outliers on the left of the simulated
trend shown in Fig.~\ref{FigBTFFE} are therefore either systems with
uncharacteristically high galaxy formation efficiency, or else systems
with unusually low dark matter content within $r_{\rm out}$ for their
virial mass.

If these galaxies are rotationally supported (such that their
  measured rotation speed is equal to their circular velocity
  $V_c(r)^2=GM(r)/r$), the latter option implies that some dark
matter is ``missing'' from the inner regions of the halo, leading to a
circular velocity at $r_{\rm out}$ that substantially underestimate
the true maximum circular velocity of the system. It would be tempting
to ascribe this result to the presence of ``cores'' in the dark halo
but we note that this would imply cores larger than the galaxy
itself. Furthermore, in that case all such outliers should have rising
rotation curves that extend out to at least $r_{\rm out}$. That is
indeed the case for the one simulated point close to the $100\%$
efficiency line (top dashed curve in Fig.~\ref{FigBTFFE}). In that
case a rotation velocity of $\sim 34.5$ km/s is measured at
$r_{\rm out}=1.5$ kpc, whereas the asymptotic maximum velocity of $62$
km/s is not reached until $r_{\rm max}\sim 18$ kpc. On the other hand,
as discussed by \citep{Oman2016}, the rotation curves of the observed
outliers are typically {\it not} rising at their outermost radius.

Inclination errors could
also potentially bring some of the outliers back to the main relation. 
This would, however, require inclination corrections of order
$15^\circ-30^\circ$, well above the estimated uncertainties in
current observations (see detailed discussion in \citealt{Oman2016}). 
If more accurate measurements of inclination angles confirm the 
current estimates, these outliers
are truly systems without explanation in our $\Lambda$CDM-based model
of galaxy formation.

%
%

\section{Summary and Conclusions}
\label{SecConc}

We use the {\small APOSTLE/EAGLE} suite of cosmological hydrodynamical
simulations to examine the scaling between baryonic mass and rotation
velocity (the ``Baryonic Tully-Fisher'' relation, BTF) of galaxies
formed in a $\Lambda$CDM universe. Our main conclusions may be
summarized as follows.

We find that the observed BTF relation is reproduced, without further
tuning, by galaxy formation simulations that reproduce the galaxy
stellar mass function and predict galaxy sizes comparable to
  observed values. This implies that: 
(i) the BTF normalization is
largely determined by matching the abundance of $L_*$ galaxies; (ii)
the slope results from the steady decline in galaxy formation
efficiency with decreasing virial mass. 
Through a
fortuitous combination of competing effects, the galaxy rotation
velocities end up being, on average, almost equal to the halo
virial velocities ($V_c \approx 1.15 V_{\rm vir}$). The scatter in the
simulated BTF relation is slightly smaller than observed, despite the
strong feedback-driven winds that regulate gas accretion and star
formation in the simulations.

The agreement between observed and simulated BTF relations does not
require, in our simulations, any special adjustment of the inner dark
matter density profile (such as the creation of a constant-density
``core'' or a substantial expansion of the central cusp) except for a
mild contraction of the halo in baryon-dominated systems.

At the very faint end (i.e., rotation velocities $\lesssim 40$ km/s)
the simulated BTF relation steepens considerably in mass as a result of the
sharp drop in galaxy formation efficiencies in low-mass halos that is required
to match the shallow faint-end of the galaxy stellar mass
function. This implies that most faint dwarfs (i.e., $M_{\rm bar}
\lesssim 10^9\, M_\odot$) should inhabit halos of very similar virial
mass and, consequently, similar maximum circular velocities.

The observed steepening of the BTF relation at the faint end is less
pronounced, and is accompanied by larger scatter than expected from
the simulations. This disagreement may be reduced by accounting for
the fact that low-mass galaxies are small: their rotation curves probe
only the rising part of the halo circular velocity profile, leading to
systematic underestimation of the asymptotic maximum circular velocity.

More difficult to reconcile with simulations is the large scatter at
the faint end of the observed BTF relation. The presence of fairly
massive galaxies with unexpectedly small rotation velocities is
particularly difficult to explain. This is because, given the small
galaxy formation efficiency of low-mass halos inherent to our model,
the baryonic mass of a galaxy places a strong lower limit on the halo
mass it inhabits, implying much higher velocities than observed. 
  Unless the interpretation of the observational data is in
  substantial error (perhaps due to severe underestimation of galaxy
  inclinations) these outliers seem to be either ``missing dark
  matter'' from the inner regions of their halos, or to have
  experienced extraordinarily efficient galaxy formation
  \citep{Oman2016}. Neither possibility is accounted for in our model,
  and therefore such systems have no counterparts in our simulations.

%
\section*{Acknowledgments}
The authors wish to thank S.-H. Oh, E. de Blok, J. Adams, M. Papastergis
and J. Bradford for
making data available in electronic form. JFN acknowledges a
Leverhulme Trust Visiting Professorship hosted at the ICC by CSF. This
work was supported by the Science and Technology Facilities Council
(grant number ST/F001166/1). RAC is a Royal Society University Research Fellow.
CSF acknowledges ERC Advanced Grant
267291 ''COSMIWAY" and JFN a Leverhulme Visiting Professor grant held
at the Institute for Computational Cosmology, Durham University. This
work used the DiRAC Data Centric system at Durham University, operated
by the Institute for Computational Cosmology on behalf of the STFC
DiRAC HPC Facility (www.dirac.ac.uk). This equipment was funded by BIS
National E-infrastructure capital grant ST/K00042X/1, STFC capital
grant ST/H008519/1, and STFC DiRAC Operations grant ST/K003267/1 and
Durham University. DiRAC is part of the National E-Infrastructure.
The research was supported in part by the European Research
Council under the European Union Seventh Framework
Programme (FP7/2007-2013) / ERC Grant agreement
278594-GasAroundGalaxies. 

\bibliography{master}
\end{document}